\documentclass{article}
\usepackage{spconf,amsmath,graphicx}
\usepackage{comment}
\usepackage{enumitem}
\usepackage{xcolor}   
\setlist{nosep, leftmargin=14pt}
\usepackage{amsfonts}
\usepackage{mwe} 
\usepackage{hyperref}

\title{Assessing the use of Diffusion models for motion artifact correction in brain MRI}
\usepackage{algorithm}
\usepackage{algpseudocode}

\name{Paolo Angella\textsuperscript{1}, Vito Paolo Pastore\textsuperscript{1*†}\thanks{*Correspondence to Vito.Paolo.Pastore@unige.it}, Matteo Santacesaria\textsuperscript{1†}\thanks{†These authors contributed equally to this work.}}
\address{\textsuperscript{1} MaLGa Center,
University of Genoa, Italy 
}
\begin{document}
\maketitle
\begin{abstract}
Magnetic Resonance Imaging generally requires long exposure times, while being sensitive to patient motion, resulting in artifacts in the acquired images, which may hinder their diagnostic relevance. 
Despite research efforts to decrease the acquisition time, and designing efficient acquisition sequences, motion artifacts are still a persistent problem, pushing toward the need for the development of automatic motion artifact correction techniques. 
Recently, diffusion models have been proposed as a solution for the task at hand. While diffusion models can produce high-quality reconstructions, they are also susceptible to hallucination, which poses risks in diagnostic applications. In this study, we critically evaluate the use of diffusion models for correcting motion artifacts in 2D brain MRI scans. Using a popular benchmark dataset, we compare a diffusion model-based approach with state-of-the-art methods consisting of Unets trained in a supervised fashion on motion-affected images to reconstruct ground truth motion-free images. 
Our findings reveal mixed results: diffusion models can produce accurate predictions or generate harmful hallucinations in this context, depending on data heterogeneity and the acquisition planes considered as input.
\end{abstract}

\section{Introduction}
Magnetic Resonance Imaging (MRI) \cite{review1} has a significant limitation: its lengthy scan time, during which patients must remain completely still. This requirement poses particular challenges for young children, claustrophobic individuals, and elderly patients. Moreover, certain movements, such as those from breathing or cardiac activity, are unavoidable.

Motion-induced artifacts in MRI have distinctive characteristics. MRI images are generated by sampling points from the Fourier transform of the target object and then reconstructing the image through an inverse Fourier transform with incomplete data. As this constitutes an inverse problem, motion-induced errors can be substantially amplified. The process of working in Fourier space produces artifacts that are fundamentally different from those in conventional radiography, where subject movement typically results in localized blur. In MRI, even minor, localized movements can affect the entire image \cite{zaitsev2015motion}.

Recent years have seen increasing efforts to use deep learning methods for retrospective motion artifact removal in MRI—techniques applied after scan completion \cite{review1}. These approaches typically rely on supervised training of deep neural networks. Notable work in this area includes \cite{al2022stacked, levac2023accelerated, oh2023annealed}. A common challenge these methods face is the reliance on simulated data for training, as real-world paired data—scans of the same patient with and without motion artifacts—are neither widely available nor simple to collect. To address this limitation, several studies \cite{liu2021learning, ghodrati2021retrospective,oh2021unpaired} have developed deep learning frameworks that perform motion correction using unpaired data.

In this work, we investigate the potential of applying 
a generative model approach based on diffusion models to the task of MRI motion artifact correction. We compare this generative method's performance against a more traditional approach using a U-Net architecture \cite{ronneberger2015u}, trained on images with simulated motion artifacts, using a widely-used benchmark dataset of brain MRI scans. This comparison aims to evaluate diffusion models' effectiveness in motion correction relative to established methods. Notably, while the U-Net architecture follows a supervised learning approach, diffusion models are unsupervised, requiring no motion-affected images during training. To thoroughly evaluate these approaches—particularly the trade-off between diffusion models' accuracy in motion-artifact correction and potential harmful hallucinations—we conduct experiments across all three available planes (sagittal, coronal, and transverse). Our results indicate that diffusion models' performance varies significantly based on data heterogeneity and the investigated scan planes, with hallucinations potentially affecting reliable clinical usage. 

\begin{figure}
\centering
\includegraphics[width=0.45\textwidth]{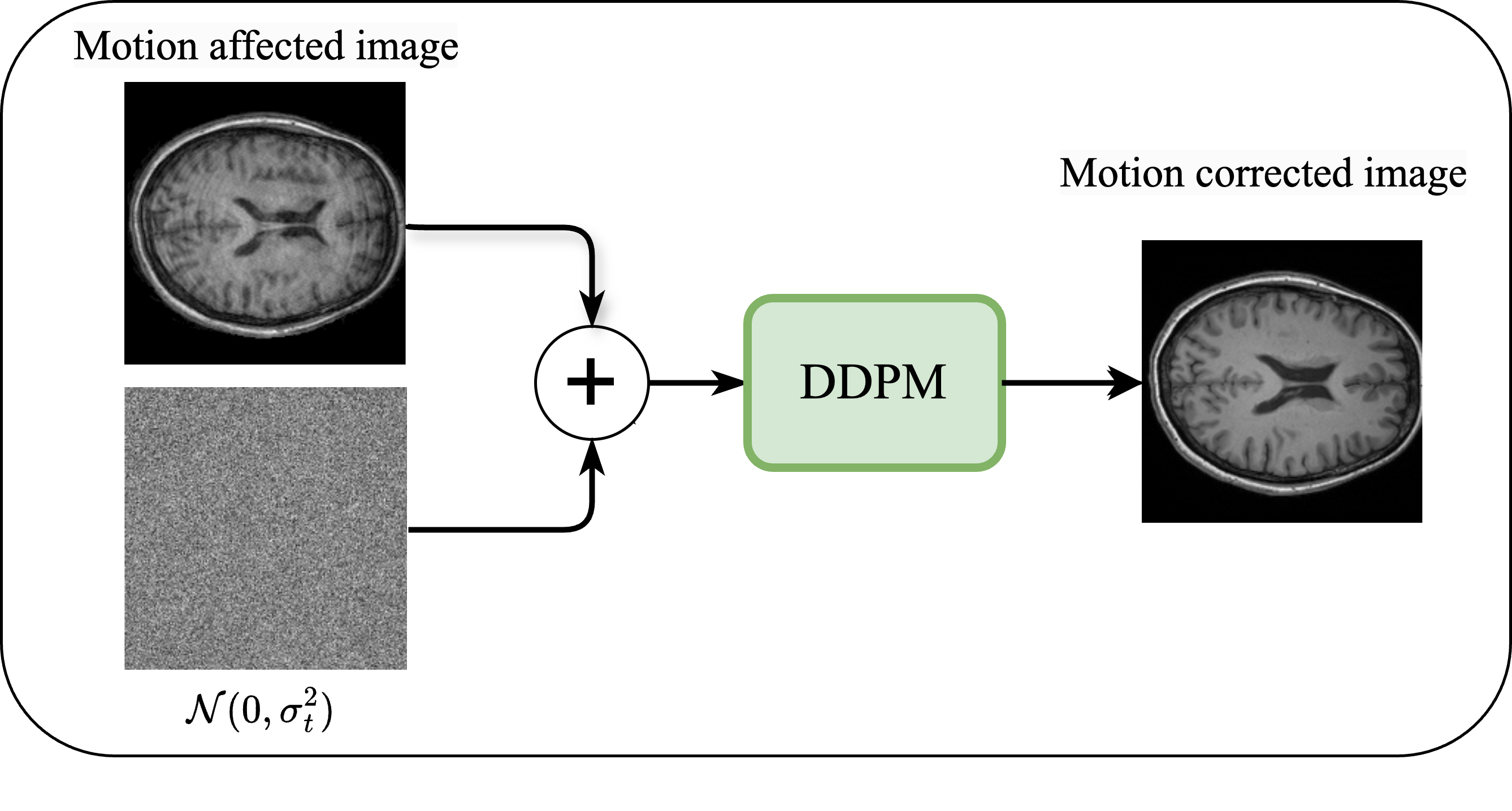}
\caption{Our implementation follows a two-phase approach. In the first phase, a diffusion model (DDPM) is trained on a dataset of motion artifact-affected images. In the second phase, the trained model is used to introduce motion artifacts into clean images, generating paired datasets. These pairs enable supervised training in the subsequent step. See Algorithm \ref{alg:motion_correction} for more details.}
\label{All}
\end{figure}

\begin{figure}
\centering
\includegraphics[width=0.475\textwidth]{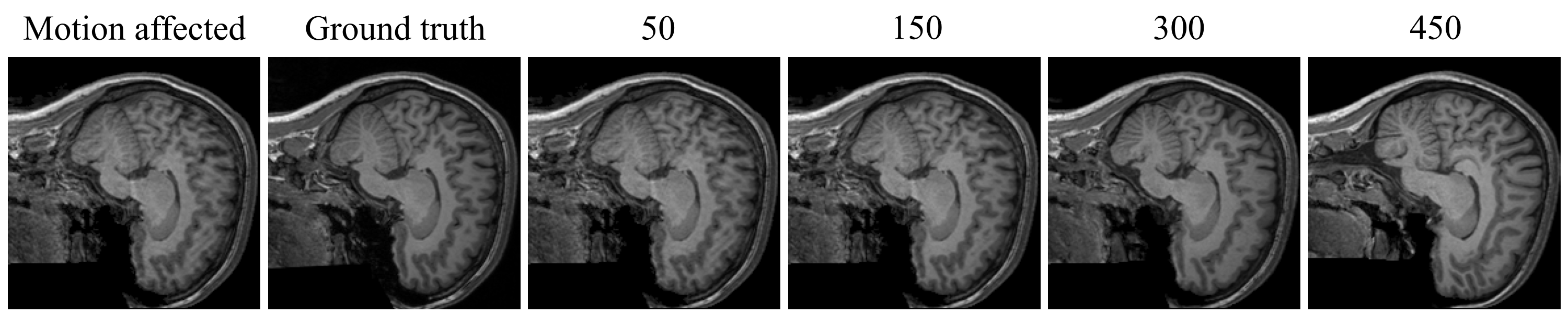}
\caption{DDPM motion artifact correction with different time steps $n$.}
\label{prog}
\end{figure}

\section{Methods}
In this work, we focused on comparing two different methods for motion correction, both of which operate under the assumption that paired data of motion-affected and motion-free images are unavailable. This assumption holds because such paired scans can only be obtained through dedicated experiments, which are not feasible in standard clinical practice. The first method involves a U-Net trained in a supervised manner using synthetically generated motion-affected images. The second method employs an unconditional diffusion model, specifically a Denoising Diffusion Probabilistic Model (DDPM) \cite{NEURIPS2020_4c5bcfec}, trained only on motion-free images, to accomplish the same task (see Fig. \ref{All} for a schematic overview).

DDPMs are generative models that generate data by reversing a diffusion process. In the forward process, Gaussian noise is gradually added to a data point $\mathbf{x}_0$ (in our case, an MRI image) over $T$ timesteps, producing a sequence of latents $\mathbf{x}_1, \dots, \mathbf{x}_T$. Each transition follows:
\begin{equation}
    q(\mathbf{x}_t | \mathbf{x}_{t-1}) = \mathcal{N}(\mathbf{x}_t; \sqrt{\alpha_t} \mathbf{x}_{t-1}, (1-\alpha_t) \mathbf{I}),
\end{equation}
where $\alpha_t$ are predefined noise variance schedule parameters. The reverse process, parameterized by a neural network, learns to denoise each $\mathbf{x}_t$ to recover $\mathbf{x}_{t-1}$:
\begin{equation}
    p_\theta(\mathbf{x}_{t-1} | \mathbf{x}_t) = \mathcal{N}(\mathbf{x}_{t-1}; \mu_\theta(\mathbf{x}_t, t), \Sigma(\mathbf{x}_t, t)).
\end{equation}
This formulation enables DDPMs to generate high-quality samples by progressive denoising starting from pure Gaussian noise.

The key challenge lies in adapting the DDPM for motion correction, as it typically generates images by starting from random noise and performing $T$ denoising steps. In our approach, described in Algorithm \ref{alg:motion_correction}, we instead begin with the motion-affected image, add Gaussian noise up to a certain step $n$ (where $n < T$), and then allow the model to denoise for the remaining $n$ steps. The choice of $n$ is crucial: if too high, the model may hallucinate details; if too low, the correction will be insufficient (see Fig. \ref{prog} for a qualitative example).

To evaluate the performance of the various approaches, we employ standard metrics for motion artifact correction assessment: the Structural Similarity Index Measure (SSIM), Normalized Mean Squared Error (NMSE), and Peak Signal-to-Noise Ratio (PSNR).

\begin{algorithm}
\caption{Diffusion-based motion correction.}
\label{alg:motion_correction}  
\begin{algorithmic}[1]
\State \textbf{input} $\mathbf{Y},n, \left\{ \alpha_t, \bar \alpha_t, \sigma_t \right\}_{t=1}^{n}, \mathbf{\epsilon}_\theta$
\State $\mathbf{z} \sim \mathcal{N}(0, \mathbf{I})$
\State $\mathbf{x}_n =  \sqrt{\bar{\alpha}_n} \mathbf{Y} + (1 - \bar{\alpha}_n) \mathbf{z}$

\For{$t = n, \dots, 1$}
\State if {$t > 1$} $\mathbf{z} \sim \mathcal{N}(0, \mathbf{I})$, else $\mathbf{z} = 0$
    \State $\mathbf{x}_{t-1} = \frac{1}{\sqrt{\alpha_t}} \left( \mathbf{x}_t - \frac{1 - \alpha_t}{\sqrt{1 - \bar{\alpha}_t}} \mathbf{\epsilon}_\theta (\mathbf{x}_t, t) \right) + \sigma_t \mathbf{z}$
\EndFor
\State \textbf{return} $\mathbf{x}_0$
\end{algorithmic}
\end{algorithm}

\section{Experiments}
\subsection{Dataset}
All experiments are conducted using the MR-ART dataset \cite{MR-ART}, which contains brain MRI scans of 144 patients, captured both while they remained still and while moving their heads to introduce motion artifacts. As mentioned earlier, having paired data like this is uncommon. Therefore, we employ data from Olsson et al. \cite{olsson2024simulating}, which includes simulated motion artifacts. The dataset consists of 3D volumes, from which sagittal, coronal, and transverse views were extracted for use in the experiments. Our preprocessing step includes normalization and registration between the motion-free and motion-affected images from MR-ART, as the volumes are often misaligned.
\subsection{Experiment Details}
All experiments are conducted on an NVIDIA A30 GPU using the PyTorch framework. The architecture of the U-Net follows the original of \cite{ronneberger2015u}, while the diffusion model is based on the architecture of \cite{ddpm-github}. Specifically, 500 timesteps were used. For DDPM training, the learning rate was set to $5 \times 10^{-5}$, and for the U-Net at $1 \times 10^{-3}$, in both cases the batch size was $6$. Early stopping was applied to all models to prevent overfitting on the training set. A subset of 30 patients was used for training, 10 for validation, and 25 for testing. This approach was adopted to ensure efficient use of computational resources while maintaining robust performance.
\subsection{Results}
We first describe our parameter-tuning procedure, and then report the obtained results on the three available anatomical planes. Our analysis compares three approaches: (1) our diffusion model approach, (2) a U-Net trained on synthetic images from \cite{olsson2024simulating}, and (3) a benchmark U-Net trained directly on paired motion-affected and motion-free images from the ground truth dataset. This last approach serves as an upper bound for performance, as it uses real paired data that would typically be unavailable in clinical settings.

\subsubsection{Parameter Tuning}
In the original MR-ART dataset, each clean image is paired with two corresponding images featuring induced motion. Therefore, for each U-Net training session, we utilize two synthetic motion-affected images for every motion-free image. In \cite{olsson2024simulating}, four sets of synthetic images were produced for each reference image. For a fair comparison, we select the best two synthetic images by training a U-Net on each possible combination (referred to as A, B, C, and D) and selecting the one yielding the best results on the validation set. All sets produce very similar outcomes, with slightly better performance for the BC set. Thus, we select this set to compare with our diffusion model-based approach.
An important parameter is the timestep at which to start the denoising process. Here, there is a trade-off between artifact removal and hallucination by the model, as shown in Fig. \ref{prog}. We selected 150 timesteps as it provided an optimal balance between reconstruction quality and the extent of hallucinations.

\begin{figure}
\centering
\includegraphics[width=0.475\textwidth]{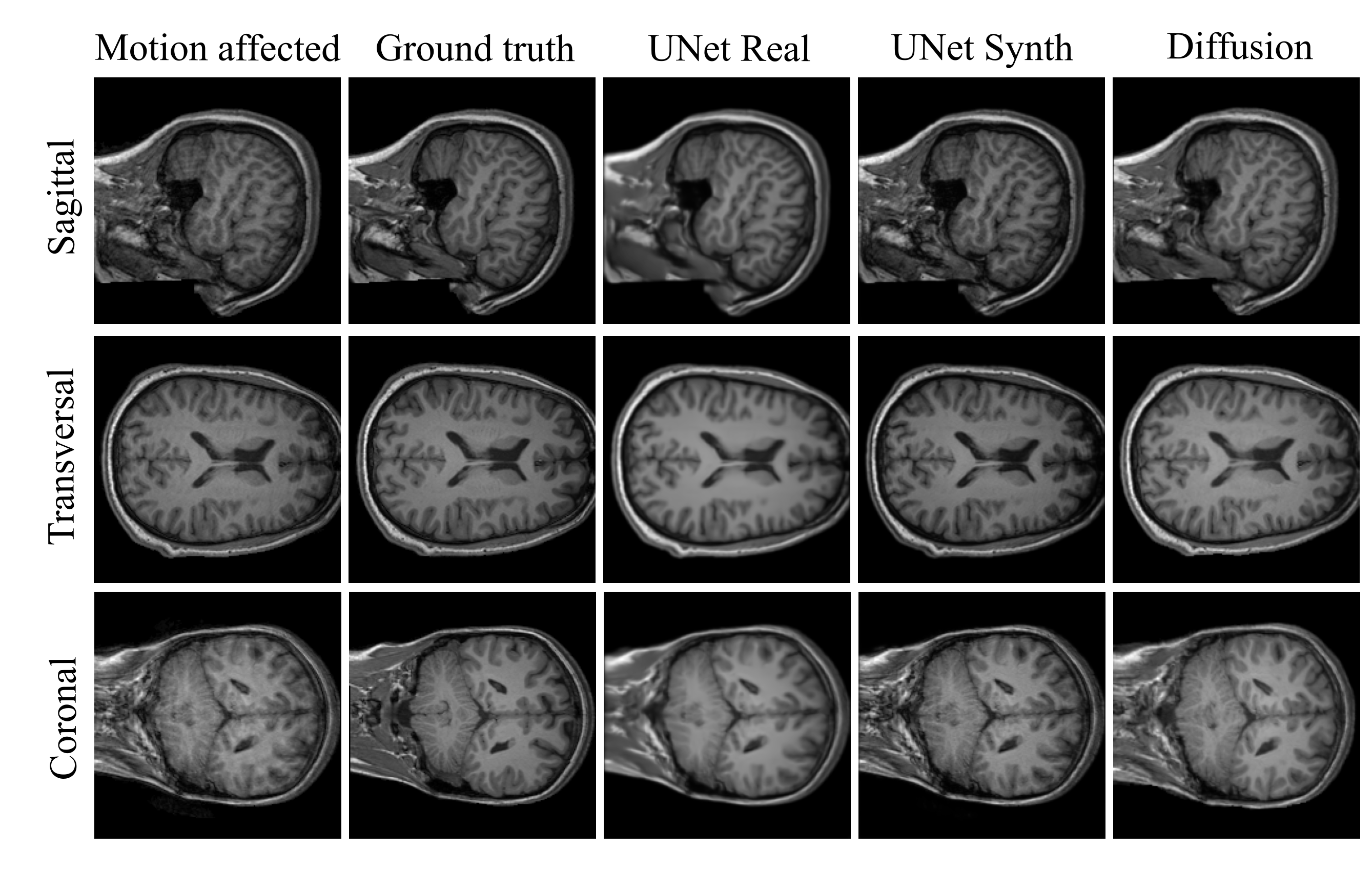}
\caption{Comparison of the different approaches on all three views.}
\label{confronto}
\end{figure}

\begin{figure}
\centering
\includegraphics[width=0.475\textwidth]{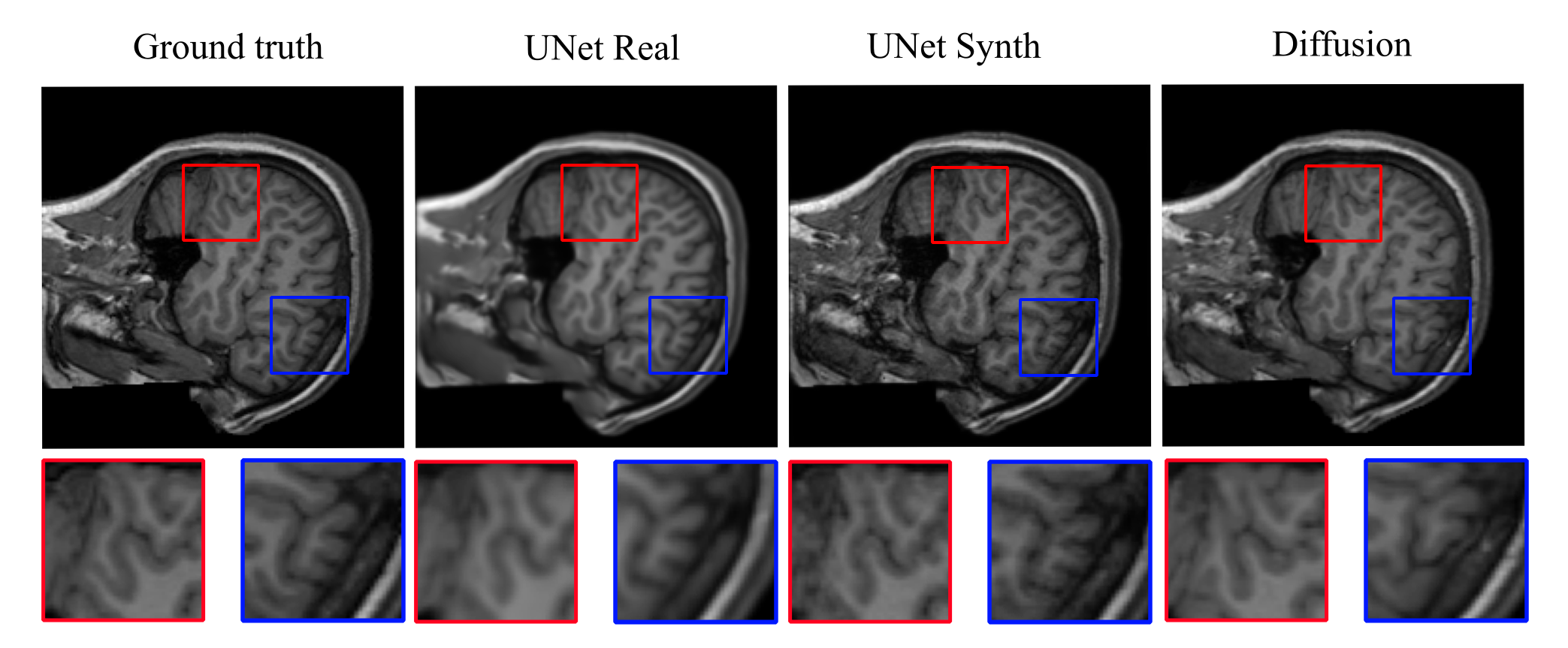}
\caption{Example of hallucination of the DDPM.}
\label{immaginie}
\end{figure}
\subsubsection{Transverse view}
\begin{table}[!ht]
\centering
\resizebox{\columnwidth}{!}{%
\begin{tabular}{llll}
\hline
Label & SSIM & NMSE & PSNR \\
\hline
UNet Real & $\textbf{0.858} \pm 0.079$ & $0.059 \pm 0.054$ & $22.9 \pm 2.7$ \\
\hline
UNet Synth & $\textit{0.844} \pm 0.088$ & $\textit{0.055} \pm 0.062$ & $\textit{24.5} \pm 2.8$ \\
Diffusion & $0.839 \pm 0.073$ & $\textbf{0.047} \pm 0.053$ & $\textbf{24.9} \pm 2.4$ \\
\hline
GAN \cite{safari23}  & $0.670$ & $0.225 $ & $21.7$ \\
Pix2Pix \cite{safari23}  & $0.640$ & $0.242$ & $21.4$ \\
Diffusion from \cite{safari23}  & $0.680$ & $0.191 $ & $21.9 $ \\
\hline
\end{tabular}}
\caption{Transverse view metrics.}
\end{table}
In the transverse view, the Diffusion model performs best for NMSE and PSNR, while the UNet achieves the highest SSIM. The UNet Synth model performs decently but slightly trails behind the Diffusion model in NMSE and PSNR, despite being close to the UNet in SSIM. See Fig. \ref{confronto} (middle) for a qualitative comparison of the motion-free images predicted with the investigated approaches. In this view, DDPM is not prone to hallucination. To better frame the obtained results in the state-of-the-art, we report the performance on the transverse view available from \cite{safari23}, exploiting popular generative models including CycleGAN \cite{zhu2017unpaired},  Pix2Pix \cite{isola2017image}, and a conditional diffusion probabilistic model. Our investigated methods outperform the available state-of-the-art results for all the evaluation metrics. Notably, the transverse is the only view employed in the literature existing on motion-artifact correction with diffusion models for the MR-ART dataset \cite{safari23}.   
\subsubsection{Sagittal view}
\begin{table}[!ht]
\centering
\begin{tabular}{llll}
\hline
Label & SSIM & NMSE & PSNR \\
\hline
UNet Real & $\textbf{0.782} \pm 0.055$ & $\textbf{0.084} \pm 0.119$ & $21.9 \pm 2.1$ \\
\hline
UNet Synth & $\textit{0.741} \pm 0.060$ & $\textit{0.098} \pm 0.124$ & $\textbf{22.2} \pm 1.9$ \\
Diffusion & $0.666 \pm 0.180$ & $0.132 \pm 0.182$ & $\textit{22.0} \pm 3.4$ \\
\hline
\end{tabular}
\caption{Sagittal view metrics.}
\end{table}
For the sagittal view, the UNet performs the best in terms of SSIM and NMSE, though the UNet Synth slightly surpasses it in PSNR. The Diffusion model performs poorly in this view, with significantly higher NMSE and a much lower SSIM (See Fig. \ref{confronto} (top) for a qualitative comparison) Furthermore, DDPM is prone to hallucination, affecting predicted motion-free images (see Fig. \ref{immaginie} for an example). 
\subsubsection{Coronal view}
\begin{table}[!ht]
\centering
\begin{tabular}{llll}
\hline
Label & SSIM & NMSE & PSNR \\
\hline
UNet Real & $\textit{0.806} \pm 0.103$ & $\textbf{0.064} \pm 0.054$ & $22.6 \pm 2.9$ \\
\hline
UNet Synth & $\textbf{0.810} \pm 0.121$ & $0.074 \pm 0.065$ & $\textbf{23.5} \pm 2.8$ \\
Diffusion & $0.765 \pm 0.108$ & $\textit{0.068} \pm 0.051$ & $\textbf{23.5} \pm 2.5$ \\
\hline
\end{tabular}
\caption{Coronal view metrics.}
\end{table}
In the coronal view, the UNet Synth performs the best overall, with the highest SSIM and PSNR, though the UNet has a lower NMSE. The Diffusion model does well in NMSE and PSNR but falls behind in SSIM compared to the UNet models, with predicted motion-free images affected by hallucination. See Fig. \ref{confronto} (bottom) for a qualitative comparison. 

\section{Conclusions}
This study critically evaluates the use of diffusion models for correcting motion artifacts in 2D brain MRI scans. Our findings reveal that while diffusion models can achieve competitive performance with supervised methods in certain scenarios, their effectiveness varies significantly with data heterogeneity and anatomical planes. The primary challenge lies in balancing reconstruction quality against the risk of hallucinations, which could be particularly problematic in diagnostic settings. Our results suggest that diffusion models hold promise for MRI motion artifact correction, but their clinical application requires careful consideration of specific use cases and potential risks. Future work should focus on developing methods to detect and mitigate hallucinations, exploring hybrid approaches that combine the strengths of both diffusion and supervised methods, and conducting clinical validation studies to assess the impact on diagnostic accuracy.

\section{Compliance with Ethical Standards}

This study was conducted retrospectively using ethically acquired publicly
available human subject data. The authors have no interests to disclose.
\section{Acknowledgments}
\label{sec:acknowledgments}
This material is supported by the Air Force Office of Scientific Research (a.n. FA8655-23-1-7083). Co-funded by Regional Problem FSE+ (point 1.2 of attach. IX of Reg. (UE) 1060/2021) and European Union - Next Generation EU.
\bibliographystyle{IEEEbib}

\end{document}